\documentclass[12pt]{iopart}
\usepackage{subcaption}
\usepackage{iopams,graphicx}
\usepackage{graphicx}
\begin{document}

\title{Uniform semiclassical approximations for one-dimensional fermionic systems}

\author{Raphael F.\ Ribeiro and Kieron Burke}

\begin{abstract}

A thorough account is given of the derivation of uniform semiclassical approximations to the particle and kinetic energy densities of N noninteracting bounded fermions in one dimension. The employed methodology allows the inclusion of non-perturbative effects via an infinite resummation of the Poisson summation formula.

\end{abstract}

\section{Introduction}
The semiclassical limit provides a variety of useful approximate solutions to
quantum mechanical problems.
Where exact numerical results are either unfeasible or provide no general insight, 
semiclassical treatments have often been used to shed light on non-perturbative effects (e.g., tunneling)
as well as to provide accurate estimates of the expectation values of
observables \cite{BM72,Ch91,BB03}.

A  quick path to the semiclassical limit of quantum mechanics is provided by the WKB approximation \cite{J24,W26, K26, B26}. However, it suffers from two important defects: i) knowledge is required of all possible solutions to the Hamilton-Jacobi equation at a given energy (or for a given time interval),
and ii) it is singular at caustics of the classical motion, where the semiclassical wave
function is also discontinuous 
as a consequence of Stokes phenomena \cite{BM72,Ch91, He13}. Therefore, WKB is of limited practical utility.
Problem i) will not concern us in this paper since our results are valid
for noninteracting fermions in one-dimension for which the classical dynamics is
trivial. Problem ii) can be excised in two different ways:
by changing the representation of the semiclassical wave function 
in the regions where it behaves pathologically (see e.g., \cite{MF01}), 
or by using uniform approximations \cite{BM72, Ch91, La37, Be69, S96}. The former method provides
local representations of the wave function which must be glued together to
generate a complete semiclassical description. Its generalization is simple \cite{MF01, Li92}.
Conversely, uniform approximations provide a global picture which is
singularity-free, but are only known in a few simple cases\cite{Ch91}.

Semiclassical uniform approximations are generally obtained in terms
of canonical functions which unfold the singularities intrinsic to primitive asymptotic treatments 
(e.g., WKB) \cite{BM72, Ch91,Be69,S96}.
The Airy function
\begin{equation} Ai(z) = \frac{1}{2\pi}\int_{-\infty}^{\infty}e^{i (t^3/3 +z t)}\mathrm{d}t \end{equation}
is the oldest and most popular of this group of special functions \cite{A38}. 
Its preponderance is attributed to the ubiquity of the fold catastrophe which arises as 
a coalescence of two non-degenerate critical points of a mapping \cite{Ch91, S96}. For example, before writing the stationary WKB wave function, the corresponding Hamilton-Jacobi equation must be solved. 
Each of its solutions is a critical point of the classical action functional. 
In one dimension, for a generic classically-allowed position and energy, there
exists two real solutions corresponding to positive and negative momentum. 
However, at a turning point there is a single zero momentum solution. As a result, the projection
of the constant energy Lagrangian submanifold on configuration space is
singular and the stationary spatial WKB wave function loses its validity \cite{A89}. 
The Airy uniform approximation is also based on the solutions to the
classical equations of motion, but it encodes this information in a way that avoids the aforementioned issue altogether \cite{BM72, Ch91, La37}.

Another attractive feature of uniform asymptotic approximations is that
they provide an understanding of the singular limits of physical theories.
In particular, uniform approximations provide an explanation of how singularities
arising in a coarse-grained description of physical phenomena (e.g., geometric optics) 
are smoothed out in a more detailed theory (e.g., wave optics) \cite{Be69}. 

The present work focuses on the semiclassical limit of sums of quantum mechanical probability densities over the lowest $N$ bound levels.  In particular, we construct
uniform semiclassical approximations to the particle and kinetic energy densities (as defined later)
of noninteracting fermionic systems in one dimension. Primitive semiclassical approximations
of limited range of validity have been obtained before for these quantities \cite{KSb65, LL75, RB08, CLEB10}.
For instance, Kohn and Sham gave region-dependent discontinuous approximations to the fermionic 
ground-state density \cite{KSb65}. Lee and Light built a similar approximation
by heuristic generalization of some properties of the linear potential Hamiltonian, but had to resort to discontinuous ad-hoc corrections from a different model to improve its accuracy \cite{LL75}.
More recently, Cangi et al. obtained a uniform approximation to the particle and kinetic energy
density, but only in the case of vanishing Dirichlet boundary conditions and Fermi energy above any critical point of the potential energy function \cite{CLEB10}.
Similarly, Roccia and Brack constructed semiclassical expressions for the density
and kinetic energy density in a classically-allowed region by applying the
stationary-phase approximation to the Gutzwiller-Van-Vleck Green function \cite{RB08}.
Notwithstanding, uniform approximations to the kinetic energy and particle densities for
the case of noninteracting fermions on $\mathbb{R}$
were lacking until our recent letter \cite{RLCE15}, which provides
contextual information, preliminary numerical analysis, and
an outline of the path towards the main results. Here
we present a detailed derivation of the uniform approximations
introduced earlier and give additional insight into their behavior. 

In Section 2 we introduce relevant definitions and establish notation. Section 3 contains the derivation of the uniform approximation to the semiclassical density, while section 4 does the same for the kinetic energy density. We conclude with open problems and future directions.  Appendices I and II provide further discussion on the smallness of higher-order terms neglected in the treatment provided in the main text.

\section{Definitions}

\subsection{Particle density and kinetic energy density}

For an isolated non-relativistic system of noninteracting $2N$ spin-1/2 fermions bound to a smooth external potential $v(x)$ with nondegenerate energy levels $E_k$, the ground-state wave function $\Psi$ can be written as the normalized antisymmetric tensor product of $N$ single-particle states (orbitals) $\psi_i(x), ~i=1,2,...,N$ satisfying $(\hat{H} \psi_i)(x) = E_i \psi_i(x)$, where $E_i < E_j,~ \forall~i < j$. The corresponding particle density $n(x;2N) \equiv n(x)$ is defined as the expectation value of the operator $\sum_{i=1}^N\delta(x-\hat{x_i})$,
\begin{equation}
n(x) = \mathrm{Tr}\left[\hat{\rho}\sum_{i=0}^{N-1} \delta\left(x-\hat{x_i}\right)\right] = 2 \sum_{i=0}^{N-1} |\psi_i(x)|^2.
\label{ndef}
\end{equation}
The kinetic energy of the same fermionic system can be obtained as the expectation value of the kinetic energy operator $\hat{T}$. But it may also be obtained by spatial integration of non-uniquely defined kinetic energy densities. In this work we denote a convenient kinetic energy density by $t(x)$ and utilize the following definition: 
\begin{equation}
t(x) = 2 \sum_{i=0}^{N-1} \psi_i^*(x)\left( \hat{T} \psi_i\right)(x).
\end{equation}
The operator identity $\hat{T} = \hat{H} - \hat{V}$ may be employed so $t(x)$ can be rewritten in a form that will find use later:
\begin{equation}
t(x) = 2 \sum_{i=0}^{N-1} \frac{p^2(x,E_i)}{2m}|\psi_i(x)|^2,
\label{tdef}
\end{equation}
where $m$ is a particle's mass, and $p^2(x,E_i)/2m = E_i - v(x)$.

Note that a nondegenerate fermionic ground state can be completely specified by its potential energy function $v(x)$ and number of particles $N$. Hence, we define the Fermi energy $E_F$ so it lies between the energy of the lowest occupied and highest unoccupied orbitals, i.e., $E_{N-1} < E_F <  E_{N}$. In this way, we may characterize a one-dimensional fermionic ground state by $v(x)$ and $E_F$. Assuming, without loss of generality, that each orbital is occupied by a single fermion of unit mass, the particle and kinetic energy densities for $N$ noninteracting bound fermions may be rewritten as:
\begin{equation}
n(x) = \sum_{i=0}^\infty |\psi_i(x)|^2 \theta \left(E_F - E_i\right),
\end{equation}
\begin{equation}
t(x) = \frac{1}{2}\sum_{i=0}^\infty |\psi_i(x)|^2 p^2(x,E_i)\theta
 \left(E_F - E_i\right),
 \label{kindef}
\end{equation}
where the spectrum of $\hat{H}$ is assumed discrete for notational purposes and $\theta(z)$ is the Heaviside step function giving 1 for $z > 0$ and $0$ for $z \leq 0$. 
\subsection{Semiclassical limit}
The semiclassical limit can be approached in various ways \cite{L07}. For instance, while it is customarily said that within a semiclassical framework $\hbar$ is taken to be arbitrarily small, this is mostly done only at a formal level. In general $\hbar$ takes its natural value while some other quantity is assumed to be large which has the same effect as that of $\hbar$ being small \cite{BM72,CLEB10, EPUR14}. 
For example, in 1973 Lieb and Simon showed the predictions of semiclassical Thomas-Fermi theory are indistinguishable from those given by the quantum theory of nonrelativistic atomic systems in the limit where nuclear charges are properly scaled \cite{LS73, L81, Sp91}.
\par Our approach towards the semiclassical limit is in the same spirit of that of Lieb and Simon. However, while the latter investigated three-dimensional Coulombic systems, we study (with a different aim) one-dimensional noninteracting models of relevance e.g., to the fields of electronic structure theory \cite{RP12} and warm-dense matter \cite{CA14}. In particular, we apply the scaling given by $\hbar \rightarrow \hbar \gamma, N \rightarrow N_\gamma = N/\gamma$ ($\gamma \in \mathbb{R}_+)$ to distinguish the dominant contributions to the particle and kinetic energy densities in the semiclassical limit, in which $\gamma \rightarrow 0$. Such scaling is discussed in more detail in Appendix I and Ref. \cite{CLEB10}.  Here we will just make the following observations: a) taking $\gamma$ to be small is equivalent to making $v(x)$ slowly-varying in the scale set by the Fermi energy; b) it is imperative that $\hbar \rightarrow 0$ and $N \rightarrow \infty$ simultaneously, for if $v(x)$ is bounded and has a single critical point, the discrete part of the corresponding Hamiltonian spectrum is finite. Therefore, $N$ cannot be taken to infinity without prior appropriate $\hbar$ rescaling. In fact, an important feature of the scaling presented is that $\gamma$ can be chosen so that any smooth binding potential allows a large finite number of bound states.  Our derivation remains valid as long as this limit is assumed. Numerical evidence for the latter argument was presented in Ref. \cite{RLCE15}.

\subsection{Semiclassical uniform approximation to the orbital wave function}

The classical limits for $n(x)$ and $t(x)$ can be derived from a plethora of methods. For instance, WKB wave functions may be used in eqs. (\ref{ndef}) and (\ref{tdef}) followed by application of the Euler-MacLaurin formula \cite{MP56} to obtain the classical limit of $n(x)$ and $t(x)$, i.e., the Thomas-Fermi density and kinetic energy density functional. Another possibility is to employ the Van Vleck-Gutzwiller Green's function \cite{VV27, Gu71} and use different representations of $n(x)$ and $t(x)$, e.g., as contour integrals in $\mathbb{C}$, to estimate their semiclassical limits by the methods of singular perturbation theory \cite{RB08}. A common feature of these treatments as well as others based on primitive WKB theory is that all of them inherit representation-dependent singularities that are intrinsic to WKB \cite{Li92}. Therefore, while such approximations provide insight into the behavior of $n(x)$ and $t(x)$ in different configuration space regions, they are both practically and theoretically unpleasant.

By definition, uniform asymptotic approximations have fractional errors tending to zero in the limit of interest for all $x \in \mathbb{R}$ \cite{BH75}. Therefore, if uniform approximations reducing to the WKB limit where appropriate (e.g., where the latter is well-defined) are employed to build semiclassical approximations for $n(x)$ and $t(x)$, then at least one source of potential singularities is eliminated. The Airy-type uniform approximation of the one-dimensional quantum wave function first obtained by Langer is particularly suitable \cite{La37}. Sufficient and necessary conditions for its validity  are that $v(x)$ must be such that the corresponding zeroes of the squared classical momentum $p^2(x,E)$ (turning points) are simple, and $p(x,E)$ is analytic everywhere except where it vanishes. From now on we assume these requirements are satisfied. Let $E$ denote the energy of a classical bound state, $\omega(E)$ the classical frequency of the periodic orbit with energy $E$, $x_-$ the l.h.s turning point for a particle with energy $E$ and $S(x,x_-,E)$ the classical action measured from $x_-$, i.e., $S(x,x_-,E) = \int_{x_-}^x p(x',E) \mathrm{d}x'$. Then, the corresponding Langer wave function can be written as:
\begin{equation}
\phi_-(x,E)=\sqrt{\frac{2m\omega(E)}{p(x,E)}}\, 
\left[\frac{3}{2} \frac{S(x,x_-,E)}{\hbar}\right]^{1/6}\, Ai\left[-\left(\frac{3}{2} \frac{S(x,x_-,E)}{\hbar}\right)^{2/3}\right],
\label{philang}
\end{equation}
where $Ai(t)$ is the Airy function evaluated at $t$ \cite{AS72}. To simplify notation, we define
\begin{equation} z(x,E) = \left[\frac{3}{2} \frac{S(x,x_-,E)}{\hbar}\right]^{2/3}.\end{equation}
An identical approximation can be made where $x_-$ is replaced by $x_+$ and the action rewritten as $S(x_+,x,E)$ so it remains positive semidefinite in the classically-allowed region. While the Airy uniform approximation was originally built for the single turning point problem it may be extended (non-uniquely) for the case where there are two such points \cite{M68}. In this work we employ the following prescription: let $x_m$ be defined such that $S(x_m,x_-,E) = S(x_+,x_m,E) = S(x_+,x_-,E)/2$. Then, for $x \leq x_m$ one may employ the left Langer wave function $\phi_-(x,E)$, while the right is used otherwise. Both will be denoted by $\phi(x,E)$ from now on.

Note that for any smooth $v(x)$ where $E$ defines a classical bound state state with two turning points, $\phi(x,E)$ is defined for all real $x$. In the classically-forbidden region, the action (and any quantities derived from it) must be analytically continued so that
$\phi(x,E)$ remains real and well-behaved. For example, for $x < x_-$, it follows that $p(x,E) = e^{i \pi/2} |p(x,E)|$, $S(x,E) = e^{3\pi i /2} |S(x,x_-,E)|$, and $z(x,E) =  (e^{3\pi i /2} 3/2 |S(x_-,x,E)|/\hbar)^{2/3}$, so 
\begin{equation}
\phi(x,E)=\sqrt{\frac{2m\omega(E)}{|p(x,E)|}}\, 
|z(x,E)|^{1/4} Ai\left(|z(x,E)|\right), x<x_-.
\end{equation}
In particular, $\phi(x,E)$ is a continuous function of $x$ across the transition region (between that which is classically-allowed and forbidden). Its behavior is oscillatory in the bulk of the classically-allowed region ($z_F(x) >>0$) \cite{AS72}. For large negative values of $z_F(x)$, i.e., for $x$ far from turning points in the classically-forbidden regions, it decays exponentially as expected for a bound finite system. Further, if the asymptotic forms of the Airy function are employed where the WKB wave function is well-defined, $\phi(x,E)$ is seen to be locally equivalent to that. 

These remarks suggest the Langer wave function provides a promising starting point for the construction of uniform approximations to the semiclassical particle and kinetic energy densities. 

\section{Uniform semiclassical approximation to the one-dimensional particle density}
\subsection{Main idea}
Our aim in this section is to obtain closed-form uniform approximations to the one-dimensional non-interacting fermionic density which respect the leading-order asymptotics of $n(x)$ everywhere in configuration space. Without loss of generality we assume orbitals are singly-occupied and the fermions have $m=1$. The external potential $v(x)$ is required to be analytic and to have non-vanishing first derivative at the turning points of all classical orbits with $E < E_F$. Under these conditions, the orbitals of  non-interacting fermionic system can be uniformly and accurately approximated by the Langer wave functions described in the previous paragraph. It follows from Eq. \ref{ndef} that the same is true for $n(x)$.  Thus, our treatment has as its starting point Eq. \ref{ndef} with Langer wave functions (Eq. \ref{philang}) employed as occupied orbitals. In what follows $E_F$ will always be chosen so that the classical action (see below) $S(E_F,x_+,x_-)$ satisfies the semiclassical quantization condition $S(E_F, x_+, x_-) = N \pi \hbar$. This choice enforces normalization of the associated Thomas-Fermi density (the leading term in any asymptotic expansion of the particle density) to $N$ particles \cite{MP56,T27,F28}. Also equivalent is to assume the Fermi level corresponds to the energy of a state with half-fractional quantum number $j=N-1/2$ in the WKB quantization condition
\begin{equation} \frac{1}{2\pi \hbar} \oint \mathrm{d}x' p[x', E(j)] = \left(j+1/2\right). \end{equation}
 For this reason every quantity evaluated at $j = N-1/2$ will be denoted by a subscript $F$. Note the above implies the Fermi energy defines a compact Lagrangian submanifold of phase space, so that no states in the continuum spectrum of $\hat{H}$ are occupied.
\par In the first step of our derivation we employ the finite Poisson summation formula \cite{Cr79}. It allows the rewriting of the particle density in a way that is amenable to a semiclassical treatment,
\begin{equation}
\sum_{j=0}^{N-1} |\psi_j|^2 =
\sum_{k=-\infty}^{\infty} \int_{\alpha-1/2}^{N+\alpha-1/2} d\lambda\,
|\psi(\lambda)|^2 e^{2\pi i k \lambda}, 
\label{fpoisson}
\end{equation}
where $-1/2<\alpha <1/2$, and $\psi(\lambda)$ fulfills the following two criteria: i) it matches $\psi_j$ when $\lambda = j$, and ii) it satisfies Dirichlet conditions in any subinterval of unit length of $(\alpha-1/2,N+\alpha-1/2)$ \cite{Cr79}. 

Using the finite Poisson summation formula with $\alpha = 0$, and the Langer wave functions for each occupied energy level we obtain for the density $n(x)$ the first approximation
\begin{equation} n(x) = \sum_{k=-\infty}^{\infty} \int_{-1/2}^{N-1/2} \mathrm{d}\lambda \frac{2 \omega(\lambda) z^{1/2}(x,\lambda)}{p(x,\lambda)}Ai^2[-z(x, \lambda)] e^{2\pi i k \lambda}. \label{npoisson}
\end{equation}
In the integrals above, physical quantities defined previously as functions of $E$ are written as functions of $\lambda$ via the mapping $E=E(\lambda)$ (e.g., $\omega(E) = \omega(E(\lambda))\equiv\omega(\lambda))$ which we assume can be well approximated by $E_{\mathrm{WKB}}(\lambda)$. Because our assumptions imply non-degeneracy of energy levels and $\mathrm{d}E/\mathrm{d}\lambda \neq 0$ for all $E\leq E_F$, the map $E(\lambda)$ is bijective in the integration interval. Note also the numerical value of each of the integrals in Eq. \ref{fpoisson} is not invariant with respect to the choice of $\alpha$. 
\par Our strategy consists of a perturbative evaluation of the integrals in Eq. \ref{npoisson}, followed by resummation of the dominant contributions to the asymptotic expansion of each integral. The terms in Eq. \ref{npoisson} where $k = 0$ and $k \neq 0$ are treated in different subsections, since their physical interpretations and asymptotic treatments are of a different nature, though, as will be seen, deeply connected. 
\subsection{Leading term}
The leading asymptotic contribution to the density in the semiclassical limit is well-known to emerge from the zeroth component of the Poisson summation formula \cite{BM72, BB03}. In other words, Thomas-Fermi theory may be obtained by approximating the summation in the definition of $n(x)$ by an integral over classical (or WKB) probability
densities \cite{BM72,CLEB10,MP56}. Thus, we expect
\begin{equation} n_0(x) =  2\int_{-1/2}^{N-1/2} \mathrm{d}\lambda \frac{\omega(\lambda)}{p(x,\lambda)} z^{1/2}(x,\lambda)Ai^2[-z(x,\lambda)],
\label{n0def}
\end{equation}
to contain the classical limit of the one-particle density $n(x)$. In what follows $x$ will be regarded as a parameter, so it will be assumed constant throughout all subsequent developments unless explicitly stated otherwise. For ease of notation we omit the spatial dependence of physical quantities at intermediate steps of the derivation. Then, upon using the identity $\hbar \omega(\lambda) \mathrm{d}\lambda =p(\lambda) \mathrm{d}p(\lambda)$, $n_0(x)$ can be rewritten in a simpler form as a Riemann-Stieltjes integral \cite{Ze95}:
\begin{equation} n_0(x) = 2\hbar^{-1} \int_{-1/2}^{N-1/2} \mathrm{d}p(\lambda) p(\lambda) f^{-1}(p) Ai^2\left[f^{-2}(p) p^2(\lambda)\right],\label{n0int0}\end{equation}
where $f(p) = f(p(\lambda)) = p(\lambda)/\sqrt{z(\lambda)} $.
Both $p(\lambda)$  and $z(\lambda)$ are of bounded variation in any compact interval of the $(x,\lambda)$ plane (see Figures 1 and 2). Additionally, the integrand is continuous in the integration domain. Therefore, the integral is well-defined.

\begin{figure}
   \begin{subfigure}{0.45\linewidth}   
      \includegraphics[width=\columnwidth,keepaspectratio]{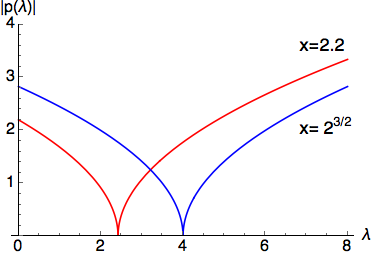}
      \subcaption{Figure 1. Generic behavior of $|p(\lambda,x)|$ for fixed $x$.}
   \end{subfigure}
   \hspace{\fill}  
   \begin{subfigure}{0.45\linewidth}
      \includegraphics[width=\columnwidth,keepaspectratio]{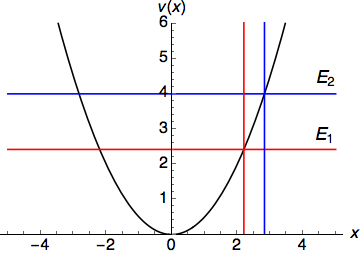}
	\subcaption{Figure 2. Typical $v(x)$; the energies $E_1$ and $E_2$ correspond to those with turning points at the two values of $x$ indicated in Figure 1.}
\end{subfigure} 
\end{figure}

If $f(p)$ were constant as is the case for the linear potential $v(x)=x$, a closed-form solution would exist for $n_0(x)$.  For well-behaved $v(x)$ we expect $f(p(\lambda))$ to be a slowly-varying function of $\lambda$. In fact, as the limit defined in Section 2.2 is approached, the variation of $f$ with respect to $\lambda$ tends to zero (see Appendix I). This suggests the following zeroth order approximation, obtained under the assumption that $f(p)$ is constant,
\begin{equation} n_0^{(0)} = \frac{p(\lambda)}{\hbar} \sqrt{z(\lambda)} \left( Ai^2[-z(\lambda)] + \frac{1}{z(\lambda)} Ai'^2[-z(\lambda)] \right) \bigg|^{\lambda = N-1/2}_{\lambda = -1/2}.\label{n0lin}\end{equation}
To extract corrections to $n_0^{(0)}$ we  take partial derivatives of the above with respect to $p$ (noting that in this case $\partial/\partial p = \partial N/\partial p ~\partial/\partial N$), change $N$ to $\lambda$, and then apply the integration operator $\int_{-1/2}^{N-1/2} \mathrm{d}p(\lambda)$ to both sides. After rearranging terms we find:
\begin{eqnarray}
\label{n0int} \fl n_0 =  \frac{p (\lambda)}{\hbar} \sqrt{z(\lambda)} \left( Ai^2[-z(\lambda)] + z^{-1}(\lambda) Ai'^2[-z(\lambda)] \right) \bigg|^{\lambda = N-1/2}_{\lambda = -1/2}+\frac{1}{\hbar}\int_{-1/2}^{N-1/2}  \mathrm{d}p(\lambda) \frac{\partial f}{\partial p} z(\lambda) Ai^2\left[-z(\lambda)\right] \nonumber \\ - \frac{1}{\hbar} \int_{-1/2}^{N-1/2} \mathrm{d}p(\lambda)\frac{\partial f}{\partial p}Ai'^2\left[-z(\lambda)\right],   
 \end{eqnarray}
where $f'(p)  = \partial f/\partial p$. The identity $\partial f/\partial p =\partial z/\partial p~ \partial f/\partial z$ allows us to rewrite the correction to $n_0^{(0)}(x)$ in a simple form:
\begin{equation}  
n_0 = n_0^{(0)} + L_{0} + \frac{1}{\hbar} \int_{-1/2}^{N-1/2} \mathrm{d}\left\{Ai[-z(\lambda)]Ai'[-z(\lambda)]\right\}\frac{\partial f}{\partial z}, \end{equation}
where $L_0$ corresponds to the first term on the r.h.s of Eq. \ref{n0int} evaluated at $\lambda = -1/2$. Further integration by parts gives:
\begin{equation}  \fl n_0 = n_0^{(0)} + \frac{1}{\hbar} \left[ \frac{\partial f}{\partial z} \bigg|_{z=z_F} Ai[-z_F]Ai'[-z_F] - \int_{-1/2}^{N-1/2} \mathrm{d}z(\lambda) \frac{\partial^2 f}{\partial z^2} Ai[-z(\lambda)]Ai'[-z(\lambda)] \right] + L(x). \label{n0int2}
 \end{equation}
where $L$ contains all previously integrated terms evaluated at $\lambda = -1/2$, i.e.,
\begin{equation} \fl \hbar L= \left[- p(\lambda) \sqrt{z_(\lambda)} \left( Ai^2[-z(\lambda)] + \frac{1}{z(\lambda)} Ai'^2[-z(\lambda)] \right) - \frac{\partial f}{\partial z} \bigg|_{z=z(\lambda)} Ai[-z({\lambda})]Ai'[-z({\lambda})] \right]\bigg |_{\lambda= - 1/2}. \label{ldef}
\end{equation}
A hint that $L(x)$ will turn out to be negligible under our assumptions is that $\lambda=-1/2$ corresponds in the WKB approximation to a classical system with zero action, i.e., $\oint \mathrm{d}x p\left[x,E(-1/2)\right] = 0$. In this case,the classical motion is supported on a minimum of $v(x)$. In Appendix I we show explicitly that both $L(x)$ and the integral in Eq. \ref{n0int2} can be safely ignored under the scaling given in Section 2.2. Hence, we find $n_0$ may be approximated under conditions of small $\hbar$ and large $N$ by:
\begin{equation} \fl  n_0 \sim  \frac{p_F}{\hbar} \sqrt{z_F}\left( Ai^2[-z_F] + z_F^{-1} Ai'^2[-z_F]\right) + \left(\frac{\omega_F}{p_F \alpha_F} - \frac{p_F}{2 \hbar z_F^{3/2}} \right) Ai[-z_F]Ai'[-z_F],
\label{n0}
\end{equation}
where $\alpha_F(x) =  z_F^{1/2}(x) \partial z(x,\lambda) /\partial \lambda |_{\lambda = N-1/2}$.  
 
Note that as $x$ approaches a turning point corresponding to the Fermi energy, $\alpha_F(x) \rightarrow 2 \hbar \omega_F z_F^{3/2}(x)/p_F^2(x)$. Thus, Eq. \ref{n0} reduces to Eq. \ref{n0lin} (minus the terms depending on $\lambda=-1/2$) in a neighborhood of each turning point. This is consistent with the assumption that there exists a region near the turning points where the potential may be linearized and where its properties become identical to those of the linear potential, a central requirement of this work. Further, use of the Airy function asymptotic expansions for large positive $z_F$ recovers the Thomas-Fermi limit for the density at leading order (see Appendix I). 

\subsection{Dominant corrections to leading term}

Let $n_1(x)$ denote the sum of the components of the Poisson summation formula with $k \neq 0$. Then, using the integral representation of $Ai^2(-z)$ \cite{LY70}, $n_1(x)$ can be expressed as:
\begin{equation}
n_1= 2 \sum_{k=-\infty}^{\infty'}\int_{-1/2}^{N-1/2} \mathrm{d}\lambda \frac{\omega(\lambda) \sqrt{z(\lambda)}}{p(\lambda)} e^{2\pi i k \lambda} \int_{\mathcal{C}}\mathrm{d}t \frac{e^{\left(t^3/12 + z(\lambda) t\right)}}{4i \pi^{3/2} \sqrt{t}},
\label{n1def}
\end{equation}
where the primed summation implies that $k \neq 0$ and the contour $\mathcal{C}$ is given in Figure 3.
\begin{figure}
\begin{subfigure}{\linewidth}   
\begin{center}
\includegraphics[scale=0.5]{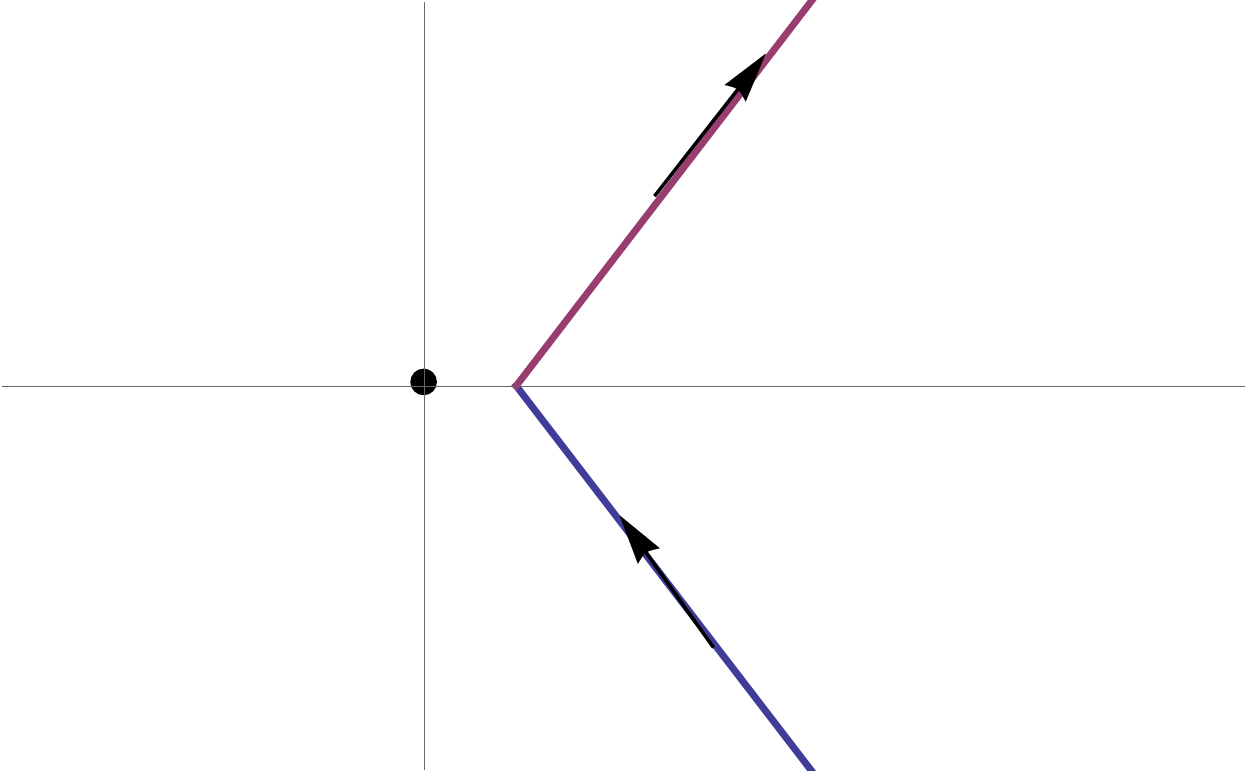}
\subcaption{Figure 3. The contour represents a closed curve starting at $\infty e^{-i\pi/3}$ and  ending at $\infty e^{i\pi/3}$. The black dot represents a branch point at $t=0$, and the branch cut is positioned at Re $t < 0$. For details, see Ref. \cite{LY70}.}
\end{center}
\end{subfigure}
\end{figure}
To obtain approximate forms for the integrals in Eq. \ref{n1def} a choice of perturbative method must be made. For that, we recourse to the following arguments. It is well-known that the semiclassical limit of the fermionic particle density is expressed in terms of quantities that depend only on the Fermi energy \cite{CLEB10,MP56, LY72,LW02}. Similarly, as consequence of the Darboux-Christoffel formula the fermionic ground-state harmonic oscillator particle and kinetic energy densities can be written exactly in terms of the lowest unoccupied orbital $\psi_N(x)$ \cite {Hu40}, e.g.,
\begin{equation} n^{\mathrm{SHO}}(x;N) = \frac{1}{2}\left(\frac{\mathrm{d}\psi_N(x)}{\mathrm{d}x}\right)^2 + \frac{1}{2}p_F^2(x) \psi_N^2(x), \end{equation}
where units were chosen so $\hbar=m=\omega=1$, single-occupation of the orbitals $\{\psi_0, ...,\psi_{N-1}\}$ was assumed, and for the harmonic oscillator $p_F(x) = \sqrt{2(N-1/2x^2)}$. These motivate our assumption that the dominant contribution to each of the integrals in $n_1(x)$ originates from a small neighborhood of $\lambda = N-1/2$ in the integration domain.

Define $F(\lambda) = 2 \pi k \lambda - i z(\lambda) t$ and $y(\lambda) = \alpha(\lambda) z^{-1/2}(\lambda)$, so  $\partial F/\partial \lambda = F'(\lambda)  = 2 \pi k - i y(\lambda) t$. Then, upon switching the integration order in Eq. \ref{n1def} we obtain:
\begin{equation} n_1 = 2 \sum_{k=-\infty}^{\infty'} \int_\mathcal{C} \mathrm{d}t \frac{e^{t^3/12}}{4i\pi^{3/2} \sqrt{t}}  \int_{-1/2}^{N-1/2} \mathrm{d}F(\lambda)  \frac{\omega (\lambda) \sqrt{z(\lambda)}}{p(\lambda) F'(\lambda)} e^{iF(\lambda)}. \end{equation} From integration by parts we find 
\begin{equation}  \fl n_1 = \frac{2\omega_F \sqrt{z_F}}{p_F} \sum_{k=-\infty}^{\infty'} (-1)^k \int_\mathcal{C} \mathrm{d}t \frac{\exp\left[ \left(t^3/12 + z_F t\right)\right]}{4i\pi^{3/2} \sqrt{t}} \frac{1}{ 2 \pi i k + z_F^{-1/2} \alpha_F  t} + R_1.  \label{n1int0} \end{equation}
The first term in the r.h.s of the above may give a useful approximation to $n_1(x)$ as long as the remainder $R_1$ is relatively small. In Appendix I, we show explicitly this is in fact the case at the semiclassical limit.

The factor $(2\pi i k + \alpha_F t/\sqrt{z_F})^{-1}$ may be expanded as a convergent geometric series in $t/(2\pi i k y_F^{-1}) $ within the disk $ |t| < t_r = |2 \pi k y_F^{-1}|$. While $y_F$ can be made arbitrarily small (but different from zero) by the scaling defined in Section 2.2, no matter how large $t_r$ is, the integration domain will contains regions where $|t| \geq t_r$.
If term-by-term integration is performed, then the resulting series will be divergent.
Similar phenomenon arises in the case of many asymptotic expansions, such as the exponential and the Stieltjes integral \cite{BH75,Di72}. The behavior of this class of asymptotic expansions is well-understood, see, e.g., \cite{ BH75,Di72}. For instance, the accuracy of estimates based on the leading term increases as the radius of convergence of the associated geometric series is enlarged. In addition, approximations obtained by the inclusion of higher-order corrections become progressively more accurate, but only until one reaches the parameter-dependent optimal truncation point where the error made by the asymptotic expansion is minimal, and beyond which the pathological behavior of the series starts to show (the magnitude of higher-order approximations increases unboundedly). 
\par For each value of $k$ in Eq. \ref{n1def}, we are only interested in the lowest-order terms. Further, as previously mentioned, the radius of convergence of the geometric series expansion of  $(2\pi i k + \alpha_F t/\sqrt{z_F})^{-1}$ is arbitrarily large in the semiclassical limit. Therefore, the pathological effects of the singularity in the integrand of Eq. \ref{n1int0} emerge only at high-order corrections for which we have no use.

Thus, we expand $(2\pi i k + \alpha_F t/\sqrt{z_F})^{-1}$ as a geometric series in $t/(2\pi i k y_F^{-1})$ and follow it by changing the order of summation and integration to encounter:
\begin{equation}\fl  n_1 \sim 2 \frac{\omega_F \sqrt{z_F}}{p_F} \sum_{j=0}^{\infty} \left( \alpha_F z_F^{-1/2} \right)^j \sum_{m=-\infty}^{\infty'} \frac{(-1)^m}{(2m\pi)^{j+1}} \int_{\mathcal{C}} \mathrm{d}t \frac{\exp\left(t^3/12 + z_F t\right)}{-4\pi^{3/2} \sqrt{t}} (it)^j. \end{equation}
The expression above may be simplified by use of the identities:
\begin{equation} \sum_{m=-\infty}^{\infty'} \frac{(-1)^m}{(2\pi m)^{j+1}} = \frac{(-1)^j 2 (2^j-1)\zeta(j+1)}{\pi^{j+1}2^{2j+1}},  ~\mbox{for $j$ odd, 0 otherwise,} \end{equation}
\begin{equation} \frac{\partial^j}{\partial z_F^j} \int_{\mathcal{C}} \mathrm{d}t \frac{\exp\left(t^3/12 + z_F t\right)}{4i\pi^{3/2} \sqrt{t}}  = \frac{\partial^{j}}{\partial z_F^j}Ai^2[-z_F], \end{equation}
where $\zeta(p)$ is the Riemann zeta function \cite{AS72}. It follows that:
\begin{equation} n_{1} \sim \frac{\omega_F \sqrt{z_F}}{p_F} \sum_{j=1}^{\infty} (-1)^{j}\ \left( \alpha_F z_F^{-1/2} \right)^{2j-1} \frac{(2^{2j-1}-1)\zeta(2j)}{\pi^{2j}4^{j-1}} \frac{\partial^{2j-1}}{\partial z_F^{2j-1}} Ai^2[-z_F]. \end{equation} 
This expression could be further simplified by using the binomial expansion for multiple derivatives of a product and a recently discovered formula for the $j$th derivative of the Airy function (so-called Airy polynomial \cite{Br13}). The end result is:
 \begin{equation}
\fl n_1 \sim \frac{\omega_F}{p_F}\sum_{q=0}^{2} \sum_{j=0}^{\infty}\left(-z_F\right)^{-3j -q} \xi_{3j+q} (\alpha_F)Ai^{(q+1)/\mathbb{Z}_3}[-z_F]Ai'^{(1-q)/\mathbb{Z}_3}[-z_F],
 \label{n1full} 
 \end{equation}
 where for $u \in \mathbb{Z}, u/\mathbb{Z}_3 = u \mathrm{~mod~}3$, and each of the $\{\xi_j(\alpha_F)\}$ is a different power series in $\alpha_F$, e.g.:
\begin{equation}
\xi_0(\alpha) = \sum_{k=1}^{\infty} \frac{(-1)^{k-1} 2 \left(2^{2k-1} -1\right) B_{2k}}{(2k)!} \alpha^{2k-1},
\label{csc}
\end{equation}
\begin{equation}
\fl \xi_1(\alpha) = \frac{7\alpha_F^3}{1440} + \frac{31\alpha_F^5}{17280} +
\frac{127 \alpha_F^7}{302400} + \frac{21127 \alpha_F^9}{27371520} + \frac{32532971 \alpha_F^{11}}{2615348736000} + \frac{548797 \alpha_F^{13}}{298896998400} +...,
\end{equation}
\begin{equation}
\fl \xi_2(\alpha) = \frac{31 \alpha_F^5}{24192}+
\frac{127\alpha_F^7}{345600} + \frac{73 \alpha_F^9}{1013760} + \frac{1414477 \alpha_F^{11}}{11887948800} + 
\frac{8191 \alpha_F^{13}}{4598415360} + \frac{16931177 \alpha_F^{15}}{67749986304000} + ..., \end{equation}
where $B_{2k}$ identifies the $2k$th Bernoulli number. 
The power series $\xi_j(\alpha)$ seem to be related to periodic functions. For example, 
\begin{equation} \xi_0(\alpha) = \csc(\alpha) - \frac{1}{\alpha}. \end{equation}
This is an important feature of the leading term in the expansion given for $n_1(x)$. Recall that $\alpha_F(x)$ (restricted to $ x_-(E_F) < x < x_m(E_F)$, or $ x_m(E_F) < x < x_+(E_F)$) is the angle-variable canonically conjugate to the Fermi action corresponding to the periodic orbit at $E_F$. Therefore, unless its image is restricted, $\alpha_F(x)$ takes an infinite number of values which differ by $\pm 2 \pi k, k \in \mathbb{Z}$. It is an interesting fact that in the approximation obtained for $n(x)$ by summing the leading terms of $n_0(x)$ and $n_1(x)$ such restriction is completely unnecessary. When the dominant term of Eq. \ref{n1full} (that with $(q,j)=(0,0)$) is added to Eq. \ref{n0}, we obtain an approximation for $n(x)$ where $\alpha_F(x)$ only occurs as the argument of a periodic function, as expected. Thus the approximation obtained by combining $n_0(x)$ and the first correction coming from $n_1(x)$ is single-valued and well-defined everywhere. Since the first few terms of some of the $\xi_j (\alpha_F)$ also appear in series expansions of trigonometric functions of $\alpha_F$ (around $\alpha_F=0$), it is expected that the connection between the corrections of $n_0(x)$ and $n_1(x)$ remains at higher-orders.
\par If only the dominant term in Eq. \ref{n1full} is retained (see Appendix I), then
\begin{equation}
n_1(x) \sim \frac{\omega_F}{p_F(x)}\xi_{0} (\alpha_F(x))Ai[-z_F(x)]Ai'[-z_F(x)].
\label{n1}
\end{equation}
The addition of the above to Eq. \ref{n0} generates the following semiclassical uniform approximation to the fermionic particle density:
\begin{equation}
\fl n^{\mathrm{sc}}(x)= \frac{p_F(x)}{\hbar}\left[ \left(\sqrt{z} \mathrm{Ai}^2(-z) +
\frac{\mathrm{Ai}^{'2}(-z)}{\sqrt{z}} \right) +
\left(\frac{\hbar\omega_F \csc[\alpha_F(x)]}{p_F^2(x)} 
- \frac{1}{2 z^{3/2}}\right) \mathrm{Ai}(-z) \mathrm{Ai}^{'}(-z)\right]_{z=z_F(x)}.
\label{nsc}
\end{equation}
\subsection{Discussion}
Equation \ref{nsc} expresses the quantum density of a fermionic system in one-dimension in terms of quantities evaluated along the complexified Lagrangian manifold defined by $H(x,p) = E_F$ where $H(x,p) = p^2/2m + v(x)$. It must be noted that while individual classical objects such as the action or momentum become purely imaginary in regions where tunneling happens, $n^{\mathrm{sc}}(x)$ remains a real positive semidefinite function for all $x \in \mathbb{R}$ as required for probability measures. In addition, $n^{\mathrm{sc}}(x)$ is continuous everywhere. It also has continuous first derivative except at the matching point (defined in Section 2.3) $x_m(E_F)$. However, in the limit of large $N$, small $\hbar$,and fixed $N\hbar$,
\begin{equation} \lim_{\epsilon \rightarrow 0} \left | \frac{\frac{\mathrm{d}n^{\mathrm{sc}}(x)}{\mathrm{d}x}\bigg|_{x=x_m +\epsilon} -  \frac{\mathrm{d}n^{\mathrm{sc}}(x)}{\mathrm{d}x}\bigg|_{x=x_m - \epsilon}}{n(x_m)} \right| \sim \frac{\hbar \omega_F}{9 N\hbar \sqrt{2[E_F-v(x_m)]}}. \end{equation}
Because $\omega_F$ and $E_F$ only depend on $N$ and $\hbar$ via the Fermi action $I_F = N \hbar$ the above indicates that in the semiclassical limit (see Appendix I) the discontinuity in the first derivative of the particle density at $x_m$ is irrelevant.
\par Higher-order corrections to the semiclassical density can in principle be included by accounting for the contributions neglected to reach Eq. \ref{nsc}, e.g., the deviation of the Langer uniform approximation from the exact single-particle states of $\hat{H}$, the remainders of the various asymptotic approximations, etc. Nonetheless we have shown before the result obtained is already of high accuracy for a variety of potentials even when the number of occupied states is $O(1)$ \cite{RLCE15}. 
\par Towards a physical interpretation of the various terms in $n^{\mathrm{sc}}(x)$, we first note the particle density can be expressed in terms of the propagator $\hat{K}(t) = e^{-i\hat{H}t/\hbar }$ in the configuration space representation, i.e.,
\begin{equation} n(x, E_F) =  \lim_{T\rightarrow \infty} \int_{-T}^T \frac{\mathrm{d}t}{t-i\gamma} e^{iE_Ft/\hbar }K(x,x,t),\end{equation}
where time-reversal invariance guarantees the Green's function is well-defined for negative propagation times \cite{LL75}.
As is well-known the propagator $K(x,x,t)$ admits an interpretation in terms of an integral over the space of closed paths based on $x$ \cite{F48}. In the semiclassical limit, $K(x,x,t)$ is expressed as a sum over amplitudes associated to topologically inequivalent closed classical orbits \cite{Li92, Gu71}. These are classified by the Morse index $\mu$. In the case of interest to this article $\mu$ is simply given by the number of times the velocity vector of a closed orbit with $x(0) = x(T) = x$ and energy $E$ changed its sign \cite{A89}. The same interpretation can be ascribed to the different components of the Poisson summation formula (see e.g., \cite{BT76, BT77}). By using the  asymptotic forms of $Ai(-z)$ and $Ai'(-z)$ in the allowed regions for classical motion at $E_F$, it is therefore unsurprising that the leading terms terms of $n^{\mathrm{sc}}(x)$ are decomposed into the two expected classes: a dominant non-oscillatory density (Thomas-Fermi) arising from the first two terms of Eq. \ref{nsc}, corresponding to the direct $t \rightarrow 0$ orbit with $\mu=0$, and an oscillatory correction obtained from the third term of Eq. \ref{nsc} which stems from the closed classical orbits with Morse index different from zero (see Appendices I and II). 

\section{Uniform semiclassical approximation to the kinetic energy density}
The kinetic energy density (KED) can be found by reasoning similar to that for the particle density. We start with the finite Poisson summation formula representation for the KED defined in Eq. \ref{tdef} with $m=1$, singly-occupied orbitals and $\psi_j$ replaced by Langer wave functions $\phi(\lambda)$ (Eq. \ref{philang}):
\begin{equation} t = \frac{1}{2}\sum_{k=-\infty}^{\infty} \int_{-1/2}^{N-1/2} \mathrm{d}\lambda ~p^2(\lambda) |\phi(\lambda)|^2 e^{2\pi i k\lambda}  .
\label{tspoisson}
\end{equation}
The dominant component can be rewritten as another Riemann-Stieltjes integral,
\begin{equation} t_0 = \frac{1}{\hbar}\int_{-1/2}^{N-1/2} \mathrm{d}p(\lambda) p^3(\lambda) f^{-1}(p) Ai^2\left[-p(\lambda)^2 f^{-2}(p) \right].\end{equation}
Assuming $f(p)$ is constant we again recover a result which is exact for a linear potential,
\begin{equation}\fl t_{0}^{(0)} = \frac{p^3(\lambda)}{6\hbar}\left( z^{1/2}(\lambda){Ai^2\left[-z(\lambda) \right]} + z^{-1/2}(\lambda) Ai'^2[-z(\lambda)] + z^{-3/2}(\lambda) Ai[-z(\lambda)]Ai'[-z(\lambda)] \right) \bigg|^{N-1/2}_{-1/2}. \end{equation}
Upon re-setting $N-1/2 \rightarrow \lambda$ in the above, following it by taking a partial derivative with respect to $p$  (for fixed $x$, but varying $\lambda$ as usual), integrating both sides from $-1/2$ to $N-1/2$, and then rearranging terms it is found that:
\begin{eqnarray} \fl t_0 =  t_{0}^{(0)} +  \frac{1}{2\hbar}\int_{-1/2}^{N-1/2}  \mathrm{d}p(\lambda) \frac{p^4(\lambda) f'(p)}{f^2(p)} Ai^2\left[-\frac{p^2(\lambda)}{f^2(p)}\right]  - \frac{1}{2\hbar} \int_{-1/2}^{N-1/2} \mathrm{d}p(\lambda) p^2(\lambda) f'(p)Ai'^2\left[-\frac{p^2(\lambda)}{f^2(p)}\right] \nonumber  \\
- \frac{1}{2\hbar} \int_{-1/2}^{N-1/2} \mathrm{d}p(\lambda) f^2(p) f'(p)  Ai\left[\frac{-p^2(\lambda)}{f^2(p)}\right] Ai'\left[\frac{-p^2(\lambda)}{f^2(p)}\right]. \label{t0full} \end{eqnarray}
Each of the remaining integrals can be evaluated perturbatively. In particular, we change variables from $p$ to $z$ so as to obtain for the first two:
\begin{eqnarray}\fl \frac{1}{2\hbar}\int_{-1/2}^{N-1/2}  \mathrm{d}p(\lambda) \frac{p^4(\lambda) f'(p)}{f^2(p)} Ai^2\left[-\frac{p^2(\lambda)}{f^2(p)}\right] - \frac{1}{2\hbar} \int_{-1/2}^{N-1/2} \mathrm{d}p(\lambda) p^2(\lambda) f'(p)Ai'^2\left[-\frac{p^2(\lambda)}{f^2(p)}\right]  \nonumber \\ \fl = \frac{1}{2\hbar}\int_{-1/2}^{N-1/2}   \mathrm{d}z(\lambda) \frac{\partial p(\lambda)}{ \partial z(\lambda)}  \frac{\partial f}{\partial p}p^2(\lambda)\left\{z(\lambda) Ai^2[-z(\lambda)] - Ai'^2[-z(\lambda)]\right\}.  \end{eqnarray}
\par As noted before, under the scaling discussed in Appendices I and II, $\frac{\partial f}{\partial z}$ is small, so the dominant term of the above can be obtained by integration by parts:
\begin{eqnarray} \fl  \frac{1}{2\hbar}\int_{-1/2}^{N-1/2}   \mathrm{d}\left\{Ai[-z(\lambda)]Ai'[-z(\lambda)]\right\} \frac{\partial f}{\partial z}p^2(\lambda) = \frac{1}{2\hbar} \frac{\partial f}{\partial z} p^2(\lambda) Ai[-z(\lambda)]Ai'[-z(\lambda)] \bigg|_{\lambda=-1/2} ^{\lambda=N-1/2}  \nonumber \\ - \frac{1}{2\hbar} \int_{-1/2}^{N-1/2} \mathrm{d}z \left[ \frac{\partial }{\partial z } \left( p^2(\lambda)\frac{\partial f}{\partial z} \right)\right] Ai[-z]Ai'[-z].\end{eqnarray} 
By the arguments discussed in Appendix I the latter term in the above equation, the last of the integrals in Eq. \ref{t0full} and all terms depending on $\lambda=-1/2$ can be neglected. Hence, the following provides the dominant component of the the defined kinetic energy density in the semiclassical limit:
\begin{eqnarray}\fl  t_0 =\frac{p_F^3\sqrt{z_F}}{6\hbar}\left[{Ai^2\left[-z_F \right]} +\frac{1}{z_F}Ai'^2[-z_F] + \left( \frac{3\hbar\omega_F}{p_F^2 \alpha_F} -\frac{1}{2z_F^2} \right) Ai[-z_F]Ai'[-z_F]\right].  \label{t0} \end{eqnarray}
The above may be rewritten in a way that makes manifest its relation to $n_0(x)$,
\begin{equation} 
t_0(x) = \frac{p_F^2(x)}{6} n_0(x) + \frac{\omega_F p_F(x)}{3\alpha_F(x)} Ai[-z_F(x)]Ai'[-z_F(x)].
\end{equation}
The higher-order terms emerging from the $k \neq 0$ components of Eq. \ref{tspoisson} are obtained by performing essentially the same calculation done for the analogous terms of $n(x)$,
\begin{equation}t_1(x) \sim \frac{1}{2}\left[ \omega_F p_F(x) \csc(\alpha_F(x)) - \frac{\omega_F p_F(x)}{\alpha_F(x)} \right]Ai[-z_F(x)]Ai'[-z_F(x)]. \end{equation}
In fact, the relationship between $n_1(x)$ and $t_1(x)$ is simple,
\begin{equation} t_1(x) = \frac{p_F^2(x)}{2}n_1(x). \label{t1} \end{equation}
Our final expression for the kinetic energy density can thus be written as:
\begin{equation} t^{\mathrm{sc}}(x) = \frac{p_F^2(x)}{6}n^{\mathrm{sc}}(x) + \frac{p_F(x)\omega_F}{3\mathrm{sin}\alpha_F(x)}Ai[z_F(x)]Ai'[-z_F(x)].  \label{tsc} \end{equation}
\par Equations \ref{t0}, \ref{t1}, and \ref{tsc} indicate a strong similarity between the uniform approximations obtained for the density and kinetic energy density. This is unsurprising from the classical point of view, for a classical distribution of particles of unit mass $\rho_{\mathrm{cl}}(x,p)$ has kinetic energy density given by $(2\pi \hbar)^{-1} \int \mathrm{d}p\rho_{\mathrm{cl}}(x,p) p^2/2 $. Thus, if the classical phase-space distribution $\rho_{\mathrm{cl}}(x,p) = 2\theta[E_F - H(x,p)]$ is employed, then the Thomas-Fermi kinetic energy density given by $p_F^3 /6\pi \hbar$ is obtained. Because the one-dimensional particle density is given in the classical limit by $n_{\mathrm{TF}} (x)= p_F(x)/\pi \hbar$, the configuration space classical kinetic energy density can be rewritten as $ n_{\mathrm{TF}}(x) p_F^2(x) /6$. This in turn explains the factor of 1/6 in Eq. \ref{tsc} as a manifestation of the classical limit of the defined quantum mechanical kinetic energy density. 
\par As a result of its simple relation to $n^{\mathrm{sc}}$ an analysis of neglected terms in the approximations made in this section is identical to those in the previous. Further discussion of this point is given in Appendix II. In Ref. \cite{RLCE15}, the accuracy of $t^{\mathrm{sc}}$ was illustrated with a Morse potential including 21 bound states.

\section{Conclusion}
We presented detailed derivations of uniform semiclassical approximations to the noninteracting fermionic ground-state density and kinetic energy density in one-dimension. Open questions naturally emerge from our treatment. They may be classified into internal or external. The former corresponds to inquiries that can be discussed within the framework developed here, whereas the latter regard applications to different systems and further generalizations.
\par A simple internal question is whether there is a general relationship between the terms in the expansions for $n_0(x)$ and $n_1(x)$ which would allow the generation of higher-order terms in $n_1$ from those of $n_0$. For example, Eq. \ref{n0} contains the factor $\alpha_F^{-1}$ which is the leading term in the Laurent series of csc$(\alpha_F)$. The remaining terms of this series are obtained from $n_1$. Because $\alpha_F$ is an angle variable and csc$(\alpha_F)$ is the simplest trigonometric function which has a simple pole at zero, $n_1$ could have been conjectured from $n_0$ without any of the calculations done in Section 3.4.
This is important because $n_0$ contains the Thomas-Fermi term which can be easily calculated for any noninteracting model, but $n_1$ is much less trivial as it includes non-perturbative effects due to an infinite number of topologically distinct closed orbits in a complexified phase space. 
Note that we do not comment here on the accuracy of our approximations for any given potential $v(x)$.
In the semiclassical limit, as described by $\gamma$-scaling in the Appendices, the derivation here given guarantees that corrections to $n^{\mathrm{sc}}(x)$ vanish pointwise (though with different rates in distinct regions of $\mathbb{R}$), i.e, can be made arbitrarily small for sufficiently
small $\gamma$.  But for a fixed $v(x)$ and number of particles, we have not explored the difficult
question of predicting, in general, the quantitative accuracy of the main results of this paper. On the other hand, 
all individual cases previously studied \cite{RLCE15}
suggest the uniform semiclassical approximations can be extremely accurate for smooth potentials satisfying the conditions previously outlined.

The behavior of various expectation values for observables depending only on local operators is also worth further study. For instance, the energy of a noninteracting fermionic system can be estimated with Eqs. \ref{nsc} and \ref{tsc} by adding the configuration space integral of $v(x) n^{\mathrm{sc}}(x)$ to that of $t^{\mathrm{sc}}(x)$. As shown in Ref. \cite{RLCE15}, a pointwise comparison 
of $n^{\mathrm{sc}}(x)$ and $t^{\mathrm{sc}}(x)$ with the corresponding TF approximations indicates the uniform approximations include all of the quantum effects missed by Thomas-Fermi theory. On the other hand,
the expectation values of configuration space observables $O(\hat{x})$ are obtained by taking the integral of $n(x)O(x)$ over all space.
In some cases, e.g., the harmonic oscillator, this averaging perfectly cancels out errors
in the Thomas-Fermi approximation, so that
TF theory provides exact results. 
The effect would obviously be reduced for any system that cannot be reasonably approximated 
by a harmonic oscillator, but it implies further study of this issue is warranted.

It would also be interesting to find alternative derivations of the uniform approximations given here. Semiclassical formulas can often be derived in more than one way, emphasizing distinct aspects of a result. For instance, Refs. \cite{CLEB10, ELCB08,ECPG14} provide three distinct derivations of 
the semiclassical approximation to $n(x)$ with $E_F > v(x) ~\forall ~x \in [0,1], n(0)=n(1)=0$. Another example is Berry and Tabor's derivation of the EBK density of states via the Poisson summation formula \cite{BT76}, followed shortly later by an alternative which employed the trace of a semiclassical action-angle variable propagator\cite{BT77}. Each different methodology brings a new light to previously obtained results. In the case of this paper, it would be particularly beneficial to have an alternative systematic construction, since our derivation employed various identities exclusive to Airy functions, making it not obvious how to extend our treatment to general systems in any finite number of dimensions. 

The simplest extensions of the formalism developed here which would still be limited to cases where classical dynamics is trivial are: a) the study of radial Coulomb problems, b) the treatment of systems with multiple potential wells, e.g., a periodic potential or a simple double well, and c) the development of uniform approximations to the density matrix. 

\par It is unclear if the obtained semiclassical uniform approximations can be systematically amended to study radial Coulomb problems. For instance, the fast variation of the Coulomb potential near its center would forbid the use of the results given here. However, only the spherically symmetric $s$-states have substantial amplitude near the origin. Therefore, it could be that except for such states (which in any case will likely require a uniform approximation not based on Airy functions \cite{La37}), our treatment remains valid.

\par Multiple potential wells in the weak coupling regime (high-energy barriers and/or large separations) would pose no challenge to the approximations here utilized,
 as to leading order in perturbation theory in the coupling constant each well can be treated independently and so the uniform approximations here presented would apply immediately as long as the Fermi energy is below all local maxima of the potential energy function. However, it is also uncertain whether there exists simple extensions of the formalism here presented which would i) account for tunneling effects between regions separated by a barrier, and ii) provide a non-singular description of the behavior of the particle density as the Fermi energy crosses critical points of the external potential $v(x)$.
 
\par The one-particle density matrix can be employed to evaluate the exchange energy. Therefore, there exists large interest in the development of semiclassical approximations to the density matrix which contain the Thomas-Fermi limit and its dominant corrections. For instance, Elliott et al. \cite{ECPG14} have recently demonstrated the low cost and high accuracy of exchange energies obtained from a semiclassical approximation to the density matrix. However, their result only applies to systems which satisfy Dirichlet boundary conditions and for which a particle with the Fermi energy would encounter no turning points in any of its possible classical paths. Hence, another direction for future research is the application of the methods here used to obtain a uniform approximation to the density matrix. However, the introduction of another degree of freedom poses additional technical difficulties, as a new set of classical singularities is introduced to the problem.

\section{Acknowledgments} We acknowledge NSF Grant No. CHE-1464795.

 \section{Appendix I - Corrections to semiclassical particle density}
 
 In the derivations of $n_0(x)$ and $n_1(x)$, we neglected two types of terms: 
remainder integrals, such as the last term of Eq. \ref{n0int2}, and
integrated quantities evaluated at the minimum of the potential well $V_0 = E(-1/2)$, e.g., Eq. \ref{ldef}. In this appendix we show that under the scaling $\hbar \rightarrow \hbar \gamma, N_\gamma \rightarrow N/\gamma$, 
the aforementioned quantities become negligible relative to those included in Eq. \ref{nsc} when 
$\gamma$ is small.

\par Before doing so, let us recall two basic facts about our choice of scaling: i) because the limit where  $\gamma \rightarrow 0$  implies $\hbar_\gamma = \gamma \hbar \rightarrow 0$, the local de Broglie wavelength associated to the Fermi energy, $|\lambda_{F\gamma}(x)| = \gamma \hbar/|p_{F\gamma}(x)|$ is almost vanishing outside a small neighborhood of $p_F^{-1}(0)$. This condition also characterizes the regions where the WKB approximation can be employed unrestrictedly \cite{BM72,Ch91}; ii) as $\gamma \rightarrow 0$, the Fermi energy is preserved, but the spacing between energy eigenvalues of the original system is reduced to enforce the condition that $N/ \gamma$ states are occupied. This can be seen by examining the behavior of the scaled quantization condition for the Fermi action,
\begin{equation} \frac{1}{2\pi \gamma \hbar} \oint p_{F,\gamma}(x) \mathrm{d}x =
\frac{N}{\gamma}, \end{equation}
whence it is seen that $p_{F\gamma}(x) = p_F(x)$, and so $E_{F\gamma} = E_F$. 
A clear example is given by the harmonic oscillator with $\omega=1$, for which $E_{F\gamma} = E_F = N \hbar$ but $E_{N+1\gamma} - E_{N\gamma} = \Delta E_\gamma = \hbar \gamma$. Hence, the number of occupied states $E_{F\gamma}/\Delta E_\gamma = N/\gamma$ as required. The analysis that follows will shed more light on some of these points.

First, note that under $\gamma$-scaling, $z_F(x) \rightarrow z_{F\gamma}(x) = \gamma^{-2/3} z_F(x)$, so \begin{equation} \fl n_{0\gamma}  = \frac{p_F}{\hbar \gamma}  \left[ \gamma^{-1/6} \sqrt{z_F} Ai^2(- z_{F\gamma})+ \frac{\gamma^{1/6}}{\sqrt{z_F}} Ai'^2(- z_{F\gamma}) +
 \gamma \left(\frac{\hbar \omega_F}{p_F^2 \alpha_F}-  \frac{1}{2\sqrt{z_F}^3} \right) Ai(-z_{F\gamma})Ai'(-z_{F\gamma}) \right]\label{n0gamma}.  \end{equation}
For any $x$ different from a turning point, $\gamma$ can be chosen small enough that
the Airy function and its first derivative are arbitrarily close to
the leading term of their asymptotic expansions. Hence, in the classically-allowed region we find,
\begin{equation} n_{0\gamma} \sim \frac{p_F}{\gamma \hbar \pi}
 - \frac{\omega_F \cos (2 S_F/\gamma\hbar)}{2\pi p_F \alpha_F} + O(\gamma)~, z_F(x) > 0, \gamma \rightarrow 0 \label{n0g}. \end{equation}
The first term is the TF contribution, while the second is the leading, spatially-oscillating correction.
Note that the oscillations become infinitely rapid in the limit. On the other hand, in the classically-forbidden region,
\begin{equation}n_{0 \gamma}\sim e^{-2 |S_F|/\hbar \gamma} \left[  \frac{\omega_F}{4\pi |p_F| |\alpha_F|} - \frac{|p_F|}{6\pi |S_F|} + O(\gamma)\right], ~ z_F(x) < 0, \gamma \rightarrow 0 \label{n0ge}\end{equation}
Here, no TF contribution ever arises, and every term vanishes exponentially with 
$1/\gamma$. 
Near a turning point of the Fermi energy the semiclassical particle density is given by:
\begin{equation} n_{0\gamma} \sim  \gamma^{-2/3} \frac{1}{\Gamma^2(1/3)}\left(\frac{2}{9 \hbar^2} \bigg| \frac{\mathrm{d}v}{\mathrm{d}x}(x_0) \bigg | \right)^{1/3}+ O(x-x_0),~ x-x_0 \rightarrow 0. \end{equation}
At this point we pause to note that the above considerations explicitly indicate that just as it occurs with other local observables, there exists no simple global expansion of the particle density in powers of $\hbar$. However, the local expansions shown above are all encapsulated by the basic result expressed in Eq. \ref{n0gamma} which will thus be used to determine negligible terms as $\gamma \rightarrow 0$ without the necessity of examining the behavior of individual terms in each region with qualitatively different behavior for the particle density.
\par We can now look at the remainder integral in Eq. \ref{n0int2}:
 \begin{equation}
  \fl
R_0 = \hbar^{-1} \int_{z_{-1/2}}^{z_F} \mathrm{d}z \frac{\partial^2 f}{\partial z^2} Ai[-z]Ai'[-z] = -\frac{1}{2\hbar}\frac{\partial^2f}{\partial z^2} \bigg|_{z_{-1/2}}^{z_F}Ai^2[-z_F] +\frac{1}{2\hbar}\int_{z_{-1/2}}^{z_F} \mathrm{d}z \frac{\partial^3 f}{\partial z^3} Ai^2[-z] .\end{equation}
Recalling that $z \in O\left(\hbar^{-2/3}\right)$, we find $R_{0\gamma}$ is $O\left(\gamma^{2/3}Ai[-z_{F,\gamma}]^2\right)$. Thus, as $\gamma\rightarrow 0$ it vanishes relative to the terms included in Eq. \ref{n0gamma}.
\par In deriving $n_0$ we also neglected
\begin{equation} \fl L(x) = \lim_{\delta \rightarrow 0} \frac{1}{\hbar}\left[- p_{\lambda} \sqrt{z(\lambda)} \left( Ai^2[-z(\lambda)] + \frac{1}{z(\lambda)} Ai'^2[-z(\lambda)] \right) - \frac{\partial f}{\partial z} \bigg|_{z(\lambda)} Ai[-z(\lambda)]Ai'[-z(\lambda)] \right]\bigg |_{\lambda=-\frac{1}{2}+ \delta}, \end{equation}
where we add to $-1/2$ a small constant $\delta \rightarrow 0$, for the Langer approximation requires turning points to be simple zeros of the classical momentum. This is not the case when $\lambda = -1/2$. In fact,
the classical region for the corresponding state is a point. Therefore, any contribution to $n_0(x)$ from this term is exponentially small and can be safely ignored.

\par Our final  approximation for $n_1(x)$ (Eq. \ref{n1}) transforms under $\gamma$ scaling as:
\begin{equation} n_{1\gamma} (x) = \frac{\omega_F}{p_F} \xi_0 (\alpha_F) Ai[-\gamma^{-2/3} z_F(x)]Ai'[-\gamma^{-2/3} z_F(x)]. \end{equation}
As expected (based on the discussion in section 3.3) $n_{1\gamma}$ is $O\left(\gamma^0\right)$, i.e., of the same order in $\gamma$ as the last two terms of Eq. \ref{n0g}. In the classically-allowed region for a particle at the Fermi energy, 
\begin{equation} n_{1\gamma} \sim - \frac{\omega_F  \xi_0 (\alpha_F)}{2\pi p_F}  \mathrm{cos}(2 S_F/\gamma \hbar)+ O(\gamma)~, z_F(x) > 0, \gamma \rightarrow 0. \label{n1g}\end{equation}
Hence, the leading correction to the Thomas-Fermi term in Eq. \ref{n0g} is of the same order as the dominant term of $n_1(x)$. Similarly, in the forbidden region for the Fermi energy,
\begin{equation} n_{1\gamma} \sim \frac{\omega_F e^{-2 |S_F|/\gamma\hbar}}{4\pi |p_F|} \left(\mathrm{csch}(|\alpha_F|)-|\alpha_F|^{-1} \right) + O(\gamma), ~ z_F(x) < 0, \gamma \rightarrow 0,\end{equation}
while near a Fermi energy turning point,
\begin{equation} n_{1\gamma} \sim  \frac{\omega_F^2}{18\Gamma(1/3)\Gamma(2/3)}\left[\frac{\mathrm{d}v}{\mathrm{d}x}(x_0)\right]^{-1} + O(x-x_0),~ x-x_0 \rightarrow 0. \end{equation}
\par We also neglected two types of terms in the derivation of $n_1(x)$. The first is 
\begin{equation} \fl R_1(x) = -2 \sum_{k=-\infty}^{\infty'} \int_\mathcal{C} \mathrm{d}t \frac{e^{t^3/12}}{4i\pi^{3/2} \sqrt{t}}  \int_{-1/2}^{N-1/2} \mathrm{d}F(\lambda) \frac{e^{iF(\lambda)}}{iF(\lambda)} \frac{\partial}{\partial \lambda} \frac{\omega(\lambda) \sqrt{z(\lambda)}}{p(\lambda) F'(\lambda)},\end{equation}
while the second consists of 
\begin{eqnarray} \fl R_2(x) \sim \frac{\omega_F}{p_F}\sum_{p=0}^{2} \sum_{j=1}^{\infty}\left(-z_F\right)^{-3j -p} \xi_{3j+p} (\alpha_F)Ai^{(1+p)/\mathbb{Z}_3}[-z_F]Ai'^{(1-p)/\mathbb{Z}_3}[-z_F] +
 \nonumber \\ \frac{\omega_F}{p_F}\sum_{p=1}^{2} \left(-z_F\right)^{-p} \xi_{p} (\alpha_F)Ai^{(1+p)/\mathbb{Z}_3}[-z_F]Ai'^{(1-p)/\mathbb{Z}_3}[-z_F]. 
 \label{R2}
\end{eqnarray}
That $R_2(x)$ is of a higher order than Eq. \ref{n1} is easy to see because $z_{F\gamma}$ is O$(\gamma^{-2/3})$ and $\alpha_{F\gamma} = \alpha_F$. Thus, all terms in Eq. \ref{R2} are relatively small compared to those in $n_1(x)$ as $\gamma \rightarrow 0$. 

In the case of $R_1(x)$ the next-order term in integration by parts will contain factors of $1/F_\lambda^{'2}$ and $1/F_\lambda^{'3}$. This will yield various power series in $x$ if the argument on section 3.3 is followed. Each contains terms in $\gamma$ that vanish relative to $n_1(x)$. 

\section{Appendix II - Higher-order terms and limits of semiclassical kinetic energy density}

From the equations defining our approximations to $t_0$ (Eq. \ref{t0}) and $t_1$ (Eq. \ref{t1}),  it is clear that except for the introduction of $p_F^2$ and rational factors, the expressions for the uniform approximation to the kinetic energy density share the same structure of those corresponding to $n_0$ and $n_1$, respectively.  Therefore, the considerations given in the previous Appendix can be applied almost verbatim to explain the smallness of the terms neglected in the derivation of $t^{\mathrm{sc}}$. In this appendix, we apply, for the sake of completeness, $\gamma$-scaling to Eq. \ref{tsc} in the regions where the kinetic energy density behaves qualitatively different. This will provide further insight into the distinguishing features of the semiclassical approximations to the particle and kinetic energy densities.

\par In the classically-allowed part of the configuration space of a particle with the Fermi energy, the kinetic energy density behaves asymptotically as:
\begin{equation} t_{\gamma} \sim \frac{p_F^3}{6 \gamma \hbar \pi}
 - \frac{\omega_F p_F \mathrm{cos} (2 S_F/\gamma\hbar)}{4\pi \mathrm{sin}(\alpha_F)}~, \gamma \rightarrow 0,~ z_F(x) > 0, \end{equation}
whereas in the evanescent and transition regions,
\begin{equation} t_{\gamma} \sim \left(\frac{2|p_F|^3}{3 |S_F|} - \frac{3 \omega_F |p_F|}{\mathrm{sinh}|\alpha_F|} \right)\frac{e^{-2|S_F|/\gamma\hbar}}{24\pi}, \gamma \rightarrow 0,~ z_F(x) < 0, \label{tev} \end{equation}
\begin{equation} t_{\gamma} \sim -\frac{|dv/dx|}{9 \Gamma(2/3) \Gamma(1/3)} + O(x-x_0), ~x-x_0 \rightarrow 0. \end{equation}
In comparison to Eq. 23 of ref. \cite{RLCE15}, Eq. \ref{tev} contains an extra factor of 2 multiplying $|p_F|^3$. The former has a typo.
\par The above equations illustrate for one last time: i) the relative dominance of the Thomas-Fermi term $p_F^3/6\pi\hbar$ in comparison to all others as $\gamma \rightarrow 0$, ii) the exponential smallness of contributions to the kinetic energy coming from regions where the Fermi energy classical motion is forbidden, and iii) the absence of a global power series expansion in any single variable which is valid for all of configuration space.

\section{References}

\bibliography{MasterOld}

\providecommand{\newblock}{}
\begin{thebibliography}{10}
\expandafter\ifx\csname url\endcsname\relax
  \def\url#1{{\tt #1}}\fi
\expandafter\ifx\csname urlprefix\endcsname\relax\def\urlprefix{URL }\fi
\providecommand{\eprint}[2][]{\url{#2}}

\bibitem{BM72}
Berry M~V and Mount K~E 1972 {\em Reports on Progress in Physics\/} {\bf 35}
  315

\bibitem{Ch91}
Child M~S 1991 {\em Semiclassical Mechanics with Molecular Applications\/}
  (Clarendon Press, Oxford)

\bibitem{BB03}
Brack M and Bhaduri R 2003 {\em Semiclassical Physics\/} Frontiers in physics
  (Westview)

\bibitem{J24}
Jeffreys H 1925 {\em Proceedings of the London Mathematical Society\/} {\bf
  s2-23} 428--436

\bibitem{W26}
Wentzel G 1926 {\em Z. Phys.\/} {\bf 38} 518

\bibitem{K26}
Kramers H 1926 {\em Z. Phys.\/} {\bf 39} 828

\bibitem{B26}
Brillouin L 1926 {\em Compt. Rend.\/} {\bf 183} 24

\bibitem{He13}
Heading J 2013 {\em An Introduction to Phase-Integral Methods\/} Dover books on
  mathematics (Dover Publications, Incorporated)

\bibitem{MF01}
Maslov V and Fedoriuk V 2001 {\em Semi-Classical Approximation in Quantum
  Mechanics\/} Mathematical Physics and Applied Mathematics (Springer
  Netherlands)

\bibitem{La37}
Langer R~E 1937 {\em Phys. Rev.\/} {\bf 51}(8) 669--676

\bibitem{Be69}
Berry M~V 1969 {\em Sci. Prog., Oxf.\/} {\bf 57} 43--64

\bibitem{S96}
Schulman L 1996 {\em Techniques and Applications of Path Integration\/} A Wiley
  interscience publication (Wiley) ISBN 9780471166108

\bibitem{Li92}
Littlejohn R 1992 {\em Journal of Statistical Physics\/} {\bf 68} 7--50 ISSN
  0022-4715

\bibitem{A38}
Airy G 1838 {\em Transactions of the Cambridge Philosophical Society\/}
  379--402

\bibitem{A89}
Arnolʹd V 1989 {\em Mathematical Methods of Classical Mechanics\/} Graduate
  texts in mathematics (Springer) ISBN 9783540968900

\bibitem{KSb65}
Kohn W and Sham L~J 1965 {\em Phys. Rev.\/} {\bf 137} A1697--A1705

\bibitem{LL75}
Lee S~Y and Light J~C 1975 {\em The Journal of Chemical Physics\/} {\bf 63}
  5274--5282

\bibitem{RB08}
Roccia J and Brack M 2008 {\em Phys. Rev. Lett.\/} {\bf 100}(20) 200408

\bibitem{CLEB10}
Cangi A, Lee D, Elliott P and Burke K 2010 {\em Phys. Rev. B\/} {\bf 81} 235128

\bibitem{RLCE15}
Ribeiro R~F, Lee D, Cangi A, Elliott P and Burke K 2015 {\em Phys. Rev.
  Lett.\/} {\bf 114}(5) 050401

\bibitem{L07}
Landsman N~P 2007 {\em Handbook of the Philosophy of Science\/} {\bf 2}
  417--553

\bibitem{EPUR14}
Engl T, Plobl P, Urbina J and Richter K 2014 {\em Theoretical Chemistry
  Accounts\/} {\bf 133} 1563 ISSN 1432-881X

\bibitem{LS73}
Lieb E~H and Simon B 1973 {\em Phys. Rev. Lett.\/} {\bf 31}(11) 681--683

\bibitem{L81}
Lieb E~H 1981 {\em Rev. Mod. Phys.\/} {\bf 53}(4) 603--641

\bibitem{Sp91}
Spruch L 1991 {\em Rev. Mod. Phys.\/} {\bf 63}(1) 151--209

\bibitem{RP12}
Ruzsinszky A and Perdew J~P 2011 {\em Computational and Theoretical
  Chemistry\/} {\bf 963} 2 -- 6 ISSN 2210-271X

\bibitem{CA14}
{Cangi} A and {Pribram-Jones} A 2014 {\em ArXiv e-prints\/} (\textit{Preprint}
  \eprint{1411.1532})

\bibitem{MP56}
March N~H and Plaskett J~S 1956 {\em Proceedings of the Royal Society of
  London. Series A. Mathematical and Physical Sciences\/} {\bf 235} 419--431

\bibitem{VV27}
Van~Vleck J~H 1928 {\em Proceedings of the National Academy of Sciences\/} {\bf
  14} 178--188

\bibitem{Gu71}
Gutzwiller M~C 1971 {\em Journal of Mathematical Physics\/} {\bf 12} 343--358

\bibitem{BH75}
Bleistein N and Handelsman R 1975 {\em Asymptotic Expansions of Integrals\/}
  (Holt, Rinehart and Winston) ISBN 9780030835964

\bibitem{AS72}
Abramowitz M and Stegun I 1965 {\em Handbook of Mathematical Functions: With
  Formulas, Graphs, and Mathematical Tables\/} Applied mathematics series
  (Dover Publications) ISBN 9780486612720

\bibitem{M68}
Miller W~H 1968 {\em The Journal of Chemical Physics\/} {\bf 48} 464--467

\bibitem{T27}
Thomas L~H 1927 {\em Math. Proc. Camb. Phil. Soc.\/} {\bf 23} 542--548

\bibitem{F28}
Fermi E 1928 {\em Zeitschrift f\"ur Physik A Hadrons and Nuclei\/} {\bf 48}(1)
  73--79 ISSN 0939-7922

\bibitem{Cr79}
Crowley B~J~B 1979 {\em Journal of Physics A: Mathematical and General\/} {\bf
  12} 1951

\bibitem{Ze95}
Zeidler E 1995 {\em Applied Functional Analysis: Applications to Mathematical
  Physics\/} ({\em Applied Mathematical Sciences\/} no v. 108) (Springer New
  York) ISBN 9780387944425

\bibitem{LY70}
Logan N~A and Yee K~S 1970 {\em SIAM Journal on Mathematical Analysis\/} {\bf
  1} 115--117

\bibitem{LY72}
Light J~C and Yuan J~M 1973 {\em The Journal of Chemical Physics\/} {\bf 58}
  660--671

\bibitem{LW02}
Littlejohn R~G and Wright P 2002 {\em Journal of Mathematical Physics\/} {\bf
  43} 4668--4680

\bibitem{Hu40}
Husimi K 1940 {\em Nippon Sugaku-Buturigakkwai Kizi Dai 3 Ki\/} {\bf 22}
  264--314 ISSN 0370-1239

\bibitem{Di72}
Dingle R~B 1973 {\em Asymptotic expansions: their derivation and
  interpretation\/} (London: Academic Press)

\bibitem{Br13}
Brychkov Y~A 2013 {\em Integral Transforms and Special Functions\/} {\bf 24}
  607--612

\bibitem{F48}
Feynman R~P 1948 {\em Rev. Modern Phys.\/} {\bf 20} 367

\bibitem{BT76}
Berry M~V and Tabor M 1976 {\em Proceedings of the Royal Society of London.
  Series A, Mathematical and Physical Sciences\/} {\bf 349} pp. 101--123

\bibitem{BT77}
Berry M~V and Tabor M 1977 {\em Journal of Physics A: Mathematical and
  General\/} {\bf 10} 371

\bibitem{ELCB08}
Elliott P, Lee D, Cangi A and Burke K 2008 {\em Phys. Rev. Lett.\/} {\bf 100}
  256406

\bibitem{ECPG14}
Elliott P, Cangi A, Pittalis S, Gross E~K~U and Burke K 2015 {\em Phys. Rev.
  A\/} {\bf 92}(2) 022513

\end{thebibliography}

\end{document}